\documentclass[aastex,superscriptaddress,showpacs,letterpaper,showkeys,preprintnumbers,altaffilletter,amssymb,amsmath,amsfonts,aps]{emulateapj}

\makeatletter

\usepackage{apjfonts}
\usepackage{color}
\usepackage{times}
\usepackage{graphicx}
\usepackage{fancyhdr}
\usepackage{float}
\usepackage{amsmath}
\usepackage{amssymb}
\usepackage{subfigure}
\usepackage[ps2pdf,bookmarksopen=true,bookmarksnumbered=true,breaklinks=true,hyperfootnotes=false,colorlinks=true]{hyperref}
\usepackage{ulem}

\shorttitle{Implications for the Origin of GRB~070201 from LIGO}
\shortauthors{Abbott et al.}

\def\@fnsymbol#1{\ifcase#1\or * \or  $+$ \or  \$ \or \#  \or \dag \or \ddag \or
$\mathsection$ \or $ \mathparagraph$ \or $\|$  \or \textordfeminine \or \textbullet
\or ** \or $++$ \or  \$\$ \or \#\#  \or \dag\dag \or \ddag\ddag \or
$\mathsection\mathsection$ \or $ \mathparagraph\mathparagraph$ \or $\|\|$  \or
\textordfeminine\textordfeminine \or \textbullet \textbullet \or *** \or $+++$
\or  \$\$\$ \or \#\#  \or \dag\dag \or \ddag\ddag \or
$\mathsection \mathsection\mathsection$ \or $ \mathparagraph
\mathparagraph\mathparagraph$ \or $\|\|\|$  \or
\textordfeminine\textordfeminine\textordfeminine \or
\textbullet\textbullet\textbullet \or \else \@ctrerr\fi}

\newcommand\checked[1]{#1}

\newcommand{\bo}{\raise-1mm\hbox{\Large$\Box$}}

\newlength{\figureWidth}
\renewcommand{\today}{\number\day\space\ifcase\month\or
  January\or February\or March\or April\or May\or June\or
  July\or August\or September\or October\or November\or December\fi
  \space\number\year}

\def\be{\begin{equation}}
\def\ee{\end{equation}}
\def\bi{\begin{itemize}}
\def\ei{\end{itemize}}
\def\ben{\begin{enumerate}}
\def\een{\end{enumerate}}

\newcommand\ligodoc{P070081-B}

 \nocite{LeeRamirezRuizReviewProgenitors2007}
 \nocite{ManouBHNSMergers2007} \nocite{OechslinJankaBNSMergersAndGW2007}

\begin{document}

\title{Implications for the Origin of GRB~070201 from LIGO Observations\\
}

\author{
B.~Abbott\altaffilmark{14},
R.~Abbott\altaffilmark{14},
R.~Adhikari\altaffilmark{14},
J.~Agresti\altaffilmark{14},
P.~Ajith\altaffilmark{2},
B.~Allen\altaffilmark{2,~51},
R.~Amin\altaffilmark{18},
S.~B.~Anderson\altaffilmark{14},
W.~G.~Anderson\altaffilmark{51},
M.~Arain\altaffilmark{39}
% ...
}
\affil{The LIGO Scientific Collaboration, http://www.ligo.org}
% ...
\altaffiltext{2}{Albert-Einstein-Institut, Max-Planck-Institut f\"ur Gravitationsphysik, D-30167 Hannover, Germany}
% ...
\altaffiltext{14}{LIGO - California Institute of Technology, Pasadena, CA  91125, USA}
% ...
\altaffiltext{18}{Louisiana State University, Baton Rouge, LA  70803, USA}
% ...
\altaffiltext{39}{University of Florida, Gainesville, FL  32611, USA}
% ...
\altaffiltext{51}{University of Wisconsin-Milwaukee, Milwaukee, WI  53201, USA}
% ...

\author{Hurley, K. C.\altaffilmark{54}}
\altaffiltext{54}{University of California-Berkeley, Space Sciences Lab, 7 Gauss Way, Berkeley, CA 94720, USA}

\begin{abstract}
\vspace*{0.2in}

We analyzed the available LIGO data coincident with GRB~070201, a short duration hard spectrum $\gamma$-ray burst
whose electromagnetically determined sky position is coincident with the
spiral arms of the Andromeda galaxy (M31). Possible progenitors of
such short hard GRBs include mergers of neutron stars or a neutron
star and black hole, or soft $\gamma$-ray repeater (SGR) flares. These
events can be accompanied by gravitational-wave emission.
No plausible gravitational wave candidates were found within a 180~s long window around the time of GRB~070201.
This result implies that a compact binary progenitor of GRB~070201, with
masses in the range $1~M_\odot < m_1 < 3~M_\odot$ and $1~M_\odot < m_2
< 40~M_\odot$, located in M31 is excluded at $> 99\%$ confidence.
Indeed, if GRB~070201 were caused by a
binary neutron star merger, we find that $D< \checked{3.5} \textrm{ Mpc}$
is excluded,
assuming random inclination, at $90\%$ confidence.
The result also implies that an unmodeled gravitational wave burst from GRB~070201 most probably
emitted less than $4.4\times 10^{-4} M_{\odot}c^2$ (7.9$\times$10$^{50}$ ergs) in any 100~ms long period within the signal region if the source was in M31 and radiated isotropically at the same frequency as LIGO's peak sensitivity ($f \approx$~150~Hz).
This upper limit does not exclude current models of SGRs at the M31 distance.

\vspace*{0.2in}
\end{abstract}

\keywords{gamma-ray bursts (\objectname{GRB\,070201}) -- gravitational waves -- compact object mergers -- soft gamma-ray repeaters }

\pacs{
04.80.Nn,
07.05.Kf,
95.85.Sz
97.60.Bw
}

\maketitle

\section{INTRODUCTION}
\label{sec:intro}

Gamma-ray bursts (GRBs) are intense flashes of $\gamma$-rays which are
observed to be isotropically distributed over the
sky~\cite[see, e.g.:][and references therein]{1973ApJ...182L..85K,ADM:Pir2005,ADM:ME2002}.
The variability of the bursts on short time scales indicates that the
sources are very compact. Combined observations, using $\gamma$-ray and x-ray satellites such as
\href{http://heasarc.gsfc.nasa.gov/docs/heasarc/missions/vela5a.html}{Vela},
\href{http://www.batse.msfc.nasa.gov/batse/}{CGRO},
\href{http://bepposax.gsfc.nasa.gov/bepposax/index.html}{BeppoSax},
\href{http://space.mit.edu/HETE/index.html}{HETE},
\href{http://www.swift.psu.edu/}{Swift},
\href{http://heasarc.gsfc.nasa.gov/docs/heasarc/missions/wind.html}{Konus-Wind}, and
\href{http://sci.esa.int/science-e/www/object/index.cfm?fobjectid=31083&fareaid_1=21&fareaid_2=21&farchive_objecttypeid=15&farchive_objectid=30995&fchoice=-1&startz=1&startpage=1}{INTEGRAL}
\cite[see][and references therein]{
1973ApJ...182L..85K,
1992Natur.355..143M, 1999ApJS..122..465P,
2000ApJS..127...59F,
1981ApSS..75...47M,
2004ApJ...611.1005G},
missions, as well as by the Interplanetary Network (IPN), with follow-up by X-ray, optical and radio telescopes of the region around GRBs,
have yielded direct observations of afterglows from $\approx$350 GRBs. In turn, host galaxies were identified for many GRBs and redshifts were determined for $\approx$125 bursts.
The redshifts indicated that GRBs are of extra-galactic origin.  Two
types of GRBs are distinguished by their characteristic duration~\cite[see][]{grbs:general:discovery:multiple-classes, 2006Natur.444.1044G} and are understood to have different origins.

Long GRBs have duration $\gtrsim 2 \textrm{ s}$. Detailed observations
of long GRBs demonstrate their association with star-forming galaxies
ranging up to a redshift of $z\approx 6.3$
\cite[see][and references therein]{2006Natur.440..184K,2006ApJ...637L..69W,2006AA...447..897J}.
Furthermore, several nearby long GRBs have been spatially
and temporally coincident with supernovae \cite[e.g.][]{Campana:2006qe,2004ApJ...609L...5M,Hjorth:2003jt,Galama:1998ea,2006ARA&A..44..507W}.

Short GRBs have duration $\lesssim 2 \textrm{ s}$.  The progenitors of
short GRBs are not so well understood.
While there are associations with distant galaxies of different types and different star formation histories, there are also powerful bursts of $\gamma$-rays from Galactic sources, such as SGR~1806$-$20 \citep{2006ApJ...640..849N,2005Natur.434.1098H,2005Natur.434.1107P}. However, statistical analyses indicate that at most 15\% of known short GRBs can be accounted for as soft $\gamma$-ray repeaters (SGRs) \citep{2006ApJ...640..849N,2007arXiv0709.4640C}.
Moreover, the spectral characteristics
and energetics of some observed short GRBs and their
afterglows seem to contradict this hypothesis in most cases
\citep{2006ApJ...640..849N}.
The current leading
hypothesis to explain most short GRBs is the merger of neutron star or
neutron star $+$ black hole binaries \cite[see for example][and
references therein]{NakarReviewArticle2006,2007ApJ...654..878B}. However, to date no
observations have definitively confirmed the association
between short GRBs and binary mergers.

Therefore, given the candidate sources, it is plausible that GRB central engines are also
strong gravitational wave (GW) emitters at frequencies accessible
to ground-based detectors like LIGO, GEO-600, and Virgo
\citep{2005PhRvD..72d2002A,0264-9381-23-19-S01,Willke:2002,1993ApJ...417L..17K,LIGO-Bursts-s3,LIGO-Inspiral-s2-bns,2004ApJ...607..384F}.
Bursts of gravitational waves are expected
to be emitted during the GRB event, with a characteristic duration comparable
to that of the associated GRB, though the amplitude and frequency spectrum
of the gravitational-wave burst are unknown.
In the case of short GRBs produced by compact binary mergers,
gravitational waves with relatively well-modeled amplitude and frequency evolution
will be emitted during the inspiral phase of the binary system, preceding the
event that produces the GRB.

\begin{figure}
\includegraphics[width=\columnwidth]{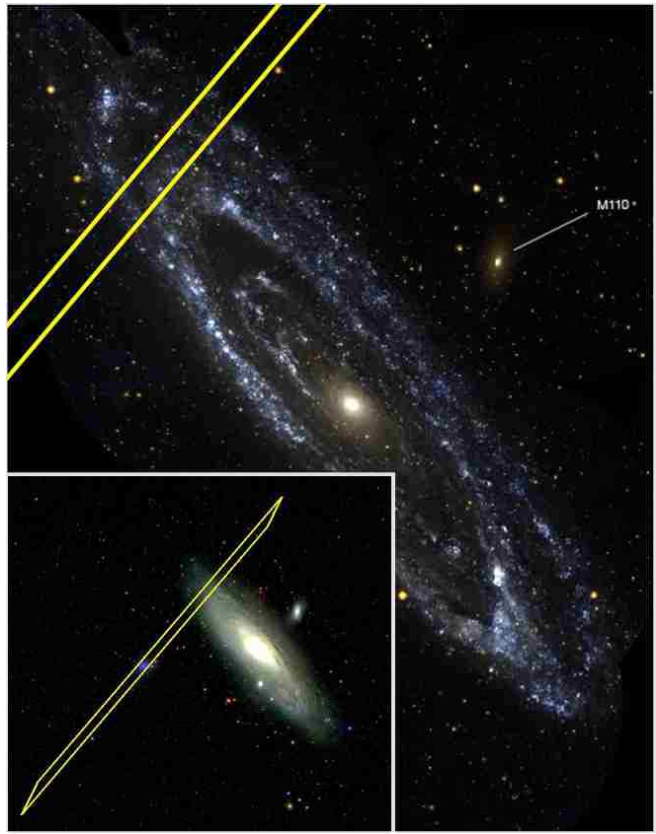}
\caption{\label{fig:M31_ErrorBox}
The IPN3~\citep{IPN} ($\gamma$-ray) error box overlaps with the spiral arms of the Andromeda galaxy (M31). The inset image shows the full error box superimposed on an SDSS~\citep{2006ApJS..162...38A,SDSS} image of M31. The main figure shows the overlap of the error box and the spiral arms of M31 in UV light~\citep{Thilker_2005ApJ}.
  }
\end{figure}

GRB~070201 was an intense,
short duration, hard spectrum GRB, which was detected and localized by 4 IPN spacecraft (Konus-Wind, INTEGRAL and MESSENGER); it
was also observed by
Swift (BAT)  but with a high-intensity background as the satellite was
entering the South
Atlantic Anomaly~\citep{2007GCN..6088....1G}.
The burst light-curve exhibited a multi-peaked pulse
with duration $\sim0.15$\,s, followed by a much weaker, softer pulse that lasted $\sim0.08$\,s. Using
early reports, Perley and Bloom~\citep{2007GCN..6091....1P}
pointed out that the initial IPN location annulus of the event
intersected the outer spiral arms of the Andromeda galaxy (M31).
The refined error box, centered $\approx$~1.1 degrees from the center of M31,
was later reported~\citep{2007GCN..6098....1P,2007GCN..6103....1H},
and it still overlaps with the spiral arms of M31 [see Figure~\ref{fig:M31_ErrorBox} and~\citep{HurleyPrivate,IPN}]. Based on the Konus-Wind observations~\citep{HurleyPrivate, 2007GCN..6094....1G},
the burst had a fluence of $1.57 (-0.21, +0.06)\times10^{-5}$~erg$\cdot$cm$^{-2}$ in the 20~keV -- 1~MeV range.

It was also pointed out~\citep{2007GCN..6094....1G} that if the burst source were actually located in M31 (at a distance of
$\simeq770$~kpc) the isotropic energy release would be $\sim10^{45}$~erg, comparable to the energy
release in giant flares of soft $\gamma$-ray repeaters: e.g., the 5$^{\rm  th}$ March 1979 event from
SGR~0526$-$66 ($\sim2\times10^{44}$~erg in the initial pulse) and the 27$^{\rm th}$ December 2004 event from
SGR~1806$-$20 ($\sim2\times10^{46}$~erg). Conversely if the event had an isotropic energy release more
typical of short hard GRBs, e.g., $\sim10^{48}-10^{52}$~erg \citep{berger-shortgrb-parameter-correlation-2007},
then it would have to be located at least $\sim30$ times further than
M31 (i.e., further than $\sim23$~Mpc).

At the  time of GRB~070201, the Hanford detectors of the Laser Interferometer Gravitational-wave Observatory
(LIGO)~\citep{LIGOS1instpaper} were stable and recording science-quality data,
while the LIGO Livingston, GEO-600, and Virgo detectors were not
taking data. The LIGO data around GRB~070201 were searched for
evidence of a gravitational wave signal from compact binary inspiral
or the central engine of the GRB itself.

A standard measure of the sensitivity of a detector to
gravitational waves is the distance to which an
optimally oriented and located double neutron star binary would produce a response
in the datastream that, when optimally filtered  for the
inspiral waves, peaks at a signal to noise ratio of 8
\citep[see, e.g.][and references therein]{LIGO-Inspiral-s2-bns}. At the time of
GRB~070201, this distance was \checked{35.7~Mpc} and \checked{15.3~Mpc}
for the Hanford 4~km and 2~km detectors, respectively.
However, the sensitivity of a detector to a
gravitational wave depends on the location of the source on the sky
and on the polarization angle of the waves. In the case of compact
binaries, it also depends on the inclination angle of the orbital
plane relative to the line of sight. At the time of GRB~070201,
the binary inspiral reach in the direction of M31
was only about $43\%$ of this maximum.
More details of the instrumental
sensitivity can be found in Sec.~\ref{sec:LIGO-Observations}.

The search for gravitational waves from a compact binary inspiral
focused on objects with masses in the ranges $1~M_\odot < m_1 < 3~M_\odot$ and $1~M_\odot < m_2 <
40~M_\odot$. The core of the
search is matched filtering, cross-correlating the data with the expected
gravitational waveform for binary inspiral  and uses
methods reported previously \citep[see, e.g][and
references therein]{LIGO-Inspiral-s2-bns}.
Uncertainties in the expected waveforms can lead to
decreased sensitivity of the search to the gravitational wave signal
from the inspiral phase; this is particularly true of systems with higher
masses and systems with substantial spin~\citep{GrandclementKalogeraVecchio:2003}.
This is accounted
for by studying the dependence of sensitivity of the search to a
variety of model waveforms based on different approximation
methods.

The search for more generic transient gravitational waves coincident with the $\gamma$-ray burst is
based on cross-correlating data from two detectors and does not make
use of a specific model for the gravitational wave signal.  This is an
appropriate method when the gravitational wave signal is not well
modeled theoretically, such as signals from the actual merger phase of a compact
binary system or the core collapse phase of a supernova event.

The remainder of the paper is organized as follows.
In Sec.~2, we discuss the LIGO detectors and the data taken
around the time of GRB~070201.  In Sec.~3, we report on the inspiral gravitational wave
search, briefly reviewing the methods and algorithms used, and concluding
with the astrophysical implications of the search for the GRB~070201 event.
In Sec.~4, we report on the search for burst-like gravitational wave signals
and present the astrophysical implications of that search.
The software used in this analysis is available in the
LIGO Scientific Collaboration's data analysis code archives \citep{lscsoft} \footnote{The search for inspiraling binaries (Sec~3.) used LAL and LALAPPS with tag s5\_grb070201\_20070731 and the
burst search (Sec.~4) used the MATAPPS package grbxcorr with tag grbxcorr\_r1}.
Since no plausible gravitational wave signal was detected above the background either in
the inspiral or the burst search, we present the astrophysical implications of
these results on the understanding of short GRBs in Sec.~5.

\section{LIGO Observations}
\label{sec:LIGO-Observations}

LIGO is comprised of three instruments at two
geographically distinct locations (a $4\textrm{ km}$ detector and a $2\textrm{
km}$ detector at Hanford Observatory, referred to as H1 and H2, and a
$4\textrm{ km}$ detector at Livingston Observatory, referred to as L1). Five science runs have been carried out to date.  GRB 070201 occurred during the most recent science run, called S5, which started on November 4th, 2005 and ended on October 1st, 2007.
All three LIGO detectors were operating at their design
sensitivity~\citep{LIGO_S5_Sensitivity} throughout the S5 run.

The LIGO detectors use suspended mirrors at the ends of
kilometer-scale, orthogonal arms to form a power-recycled Michelson interferometer
with Fabry-Perot cavities.
A gravitational wave induces a time-dependent strain $h(t)$ on the
detector.  While acquiring scientific data, feedback to the
mirror positions and to the laser frequency keeps the optical cavities
near resonance, so that interference in the light from the two arms
recombining at the beam splitter depends on the difference between the
lengths of the two arms. A photodiode senses the light, and a
digitized signal is recorded at a sampling rate of 16384~Hz. The data are calibrated and converted into a strain time series.

The LIGO detectors have a sensitive frequency band extending from $\sim$40~Hz to $\sim$2000~Hz, with the maximum sensitivity at ${\approx}150$~Hz,
which is limited at low frequencies by seismic noise and at high
frequencies by laser shot noise. In addition, environmental
disturbances, control systems noise, and other well understood noise sources
result in a non-stationary and non-Gaussian background.

\subsection{LIGO observations coincident with GRB~070201}

At the time of the GRB trigger, both LIGO Hanford detectors were
stable and recording science quality data. These detectors had been in science mode for more than
14 hours before the GRB trigger, and stayed in science mode for more
than 8 hours after the GRB trigger, providing ample data for background studies.

An asymmetric 180~s \textit{on-source} segment, $-$120$/+$60~s
about the GRB trigger time, was searched for gravitational-wave
signals. This choice~\citep{2005PhRvD..72d2002A,S2-S3-S4_GRB} is conservative enough to accommodate inspiral type signals, trigger time ambiguities, and theoretical uncertainties.
We also implicitly assume that the propagation speed of the gravitational waves is the speed of light.
The significance of candidate events was evaluated using
studies covering several hours of \textit{off-source} data from the
same science mode stretch outside of, but near to, the on-source segment.

The ideal response of a detector to an incident
gravitational wave is a weighted combination of the two underlying
gravitational wave polarizations denoted by $h_+(t)$
and $h_\times(t)$:
\begin{equation}
h(t) = F_+(\theta,\phi,\psi)\, h_+(t) +
F_{\times}(\theta,\phi,\psi)\, h_\times(t) \,.
\label{eq:hresponse}
\end{equation}
The dimensionless weighting amplitudes, or \textit{antenna factors}, $F_+$ and
$F_\times$ depend on the position $(\theta,\phi)$ of the source relative to the detector and the
gravitational wave polarization angle $\psi$.
For the location of GRB~070201, the root-mean-square
(RMS) antenna factor, $F_{\mathrm{RMS}}$, for both colocated and coaligned Hanford detectors was
\begin{eqnarray}
F_{\mathrm{RMS}} &=& \sqrt{ (F_{+}^2 + F_{\times}^2)/2 } = 0.43 / \sqrt{2} = 0.304 \, ,
\end{eqnarray}
a combination which does not depend on the polarization angle $\psi$.
Despite the sub-optimal location of GRB~070201 for the
LIGO Hanford detectors, they still had significant sensitivity for the
polarization state compatible with the detector.

\subsection{Data quality for the times surrounding the GRB~070201 trigger}

A suite of data quality tests were applied to LIGO
data. No anomalous behavior was found in either instrument at the time of
GRB~070201. On the other hand, a number of data quality issues were
identified in the off-source time used for background estimation (which amounted to $60084 \textrm{ s}$, or $16.7 \textrm{ hr}$).
Triggers were excluded from 530~s of coincident, off-source
data so identified, or 0.9\% of the off-source time.

Overflows in digital signals used in the feedback control systems were responsible
for 29~s in H1 and 29~s in H2 of excluded time.
Seismic noise in the 3--10~Hz band
known to produce false alarms in H1 was used to veto 160~s of
data. Disturbances that produced a loss in power in the H2 detector arm cavities larger than 4\% were also vetoed,
amounting to 163~s, which include 11~s when there were overflows in H2.
No such fluctuations in arm power were observed in H1.

Additionally, in the search for a compact binary progenitor, there were losses in off-source live-time due to quantization on 180~s intervals.  Each of these intervals was intended to be a trial treated the same as the central, on-source interval, for use in background determination.

For the burst analysis, three hours of data were used for the purpose of background estimation. The same data quality flags were considered as were used in the inspiral search but, due to the shorter length of the background used, only one data quality flag (an overflow in the H2 signal) was applied vetoing one of the 180~s segments in the three-hour background period.

Finally, 160 s of the off-source time were excluded from this data analysis, as it contained simulated signals. These were injected intentionally into the hardware at predetermined times to validate the detector response and signal detection algorithms.

\section{Search for gravitational waves from a compact binary
progenitor}
\label{sec:inspiral}

A number of searches for gravitational waves from compact binaries
have been completed on the LIGO data~\citep{LIGO-Inspiral-s2-bns,LIGOS2bbh,LIGOS2macho,LIGOS3S4all}.  Similar search methods were applied to the on-source time around
GRB~070201~\citep{LIGOS3S4all}. In this section we briefly describe those
methods, report the results of the search, and discuss their interpretation.

\subsection{Search Method}
\label{subsec:inspiral-method}

The core of the inspiral search involves correlating the LIGO data
against the theoretical waveforms expected from compact binary
coalescence, i.e., matched filtering the data~\citep{wainstein:1962}. The
gravitational waves from the inspiral phase, when the binary orbit
tightens under gravitational-wave emission prior to merger, are
accurately modeled in the band of LIGO sensitivity for a wide range
of binary masses~\citep{Blanchet:2006av}. The expected
gravitational-wave signal, as measured by LIGO, depends on the masses
and spins of the binary elements, as well as the spatial location,
inclination and orientation of the orbital axis. In general, the
power of matched filtering depends most sensitively on accurately
tracking the phase evolution of the signal. The phasing of compact
binary inspiral signals depends on the masses and spins, the time of
merger, and an overall phase.  In a search for gravitational waves
from compact binaries, one therefore uses a discrete set of
\textit{template waveforms} against which the data are correlated.

In this search, we adopt template waveforms which span a
two-dimensional parameter space (one for each component mass) such that
the maximum loss in signal-to-noise (SNR) for a binary with negligible
spins would be $3\%$.  While the spin is ignored in the template
waveforms, we show below that the search is still sensitive to
binaries with most physically reasonable spin orientations and
magnitudes with only moderate loss in sensitivity.  To generate a
GRB, at least one of the objects in a compact binary
must be a material object, probably a neutron star, while the second
object must either be a neutron star or a stellar mass black hole with
low enough mass~\citep{Vallisneri:1999nq,Rantsiou:2007ct} to cause disruption of the neutron star before it
is swallowed by the hole.  The mass-parameter space covered by the
templates is therefore $1~M_\odot < m_1 < 3~M_\odot$ and $1~M_\odot <
m_2 < 40~M_\odot$. The number of template waveforms required to achieve
this coverage depends on the detector noise curve; at the time of the
GRB, 7171 and 5417 templates were required in H1 and H2, respectively.

The data from each of the LIGO instruments are filtered through the
bank of templates. If the matched filter signal-to-noise exceeds a
threshold $\rho^\ast$, the template masses and the time of the maximum
signal-to-noise are recorded. For a given template, threshold
crossings are clustered using a sliding window equal to the duration
of the template as explained in \citep{findchirppaper}.
For each trigger identified in
this way, the coalescence phase and the effective distance---the
distance at which an optimally oriented and located binary would give
the observed signal-to-noise assuming masses to be those of the
template---are also
computed. Triggers identified in each instrument are further required
to be coincident in the time and mass parameters between the two
operating instruments taking into account the correlations between
those parameters. This significantly reduces the number of
background triggers that arise from matched filtering in each
instrument independently.  Because H1 was more sensitive than H2,
two different thresholds were used in the matched filtering step:
$\rho^\ast = 5.5$ in H1 and $\rho^\ast=4.0$ in H2.  This choice takes
advantage of the better sensitivity in H1 while still using H2 to
reduce the rate of accidentals.

To further reduce the background, two signal-based tests are
applied to the data.  First, a $\chi^2$ statistic~\citep{Allen:2004}, which
measures the quality of the match between the data and the template,
is computed; triggers with large $\chi^2$ are discarded.
Second, the $r^2$ veto \citep{Rodriguez:2007}, which looks at the time the
$\chi^2$ statistic stays above a threshold, is applied.

The SNR and $\chi^2$ from a single detector are combined
into an effective SNR~\citep{LIGOS3S4all}.
The effective SNRs from the two detectors are then added in
quadrature to form a single quantity $\rho_\mathrm{eff}^2$
which provides good separation between signal candidate events and
background.
The final list of coincident triggers are then
called \textit{candidate events}.

\subsection{Background and Results}

Gravitational-wave detectors are susceptible to many sources of
environmental and intrinsic noise. These sources often result in
non-Gaussian and non-stationary noise backgrounds. In the
case of H1 and H2, which share the same vacuum enclosure, these
backgrounds are correlated. To estimate the
background in this search, an equal number of 180~s
off-source segments were selected to the past and future of the
GRB trigger. All of the data, including the on-source
segment, were analyzed using the methods described above.
Triggers arising from the on-source segment were then removed, as were triggers within bad quality segments, leading to an estimate of the number of accidental triggers per 180~s segment.
A total off-source time of 56340~s was
analyzed, corresponding to 313 trials of 180~s. The
mean rate of coincidences was 2.4 per 180~s segment.

\begin{figure}[h]
\begin{center}
\includegraphics[width=\columnwidth]{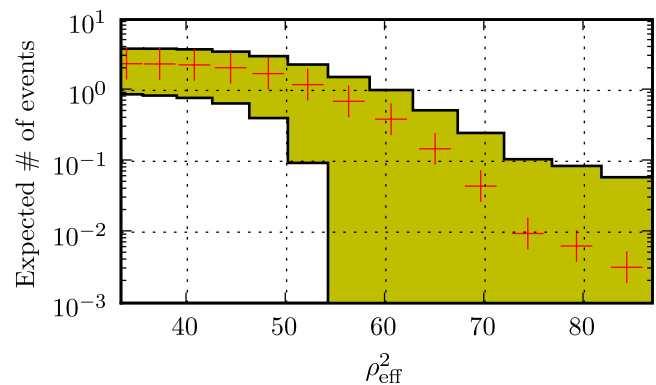}
\caption{\label{f:resultHist}
A cumulative histogram of the
 expected number of background triggers in 180~s based on the analysis of the
off-source times (pluses) as a function of the effective signal-to-noise
ratio~\citep{LIGOS3S4all}.  The shaded region indicates the $1\sigma$
variation in the background estimate observed in the off-source
times.
}
\end{center}
\end{figure}

Figure~\ref{f:resultHist} shows the expected number of
coincidences above each $\rho_{\mathrm{eff}}^2$ value in 180~s based on the analysis of the
off-source times~\citep{LIGOS3S4all}.   No candidates were observed in the
on-source time. Therefore, no plausible gravitational wave signals from
compact binary coalescence were identified around the time of
GRB~070201.

\subsection{Astrophysical Interpretation}

The observations reported here can be used to constrain the distance
to the GRB assuming it was caused by a compact binary merger. With similar considerations, one can also
evaluate the probability that a compact binary progenitor at the distance of M31 was responsible for GRB 070201.

We discover these bounds by computing the likelihood of our observation, namely the probability that no signal would be observed in the on-source time,
given the presence of a compact binary progenitor with various parameters.
Denote the gravitational-wave signal by $h(t; m_2, D, \vec{\mu})$
where $m_2$ is the mass of the companion, $D$ is the
physical distance to the binary, and $\vec{\mu} = \{ m_1, \vec{s}_1,
\vec{s}_2, \iota, \Phi_0, t_0\}$ is the mass of the neutron star, the
spins, the inclination, the coalescence phase, and the coalescence
time. The probability of interest is then
\begin{equation}
\label{inspiralProb}
p[0 | h(t; m_2, D)]
= \int \! p(\vec{\mu}) \, p[ 0 | \, h(t; m_2, D, \vec{\mu})] \, d\vec{\mu}
\end{equation}
where the nuisance parameters $\vec{\mu}$ are integrated over some
prior distribution $p(\vec{\mu})$.  This integration was performed by
injecting simulated signals into the data streams of both detectors
according to the desired prior distribution, and evaluating the
efficiency for recovering those injections as candidate events (as
described in Sec.~\ref{subsec:inspiral-method}), as a function of
$m_2$ and $D$. We choose uniform priors over $m_1$ ($1 M_\odot < m_1 < 3 M_\odot$), $\Phi_0$, $t_0$ and the polarization angle; the priors for spin and inclination $\iota$ are discussed below.

Astrophysical black holes are expected to have substantial spin. The maximum
allowed by accretion spin-up of the hole is $(a/M) = (cS/GM^2) <
0.9982$~\citep{Thorne:1974ve} in units of the Kerr spin parameter ($S$
is the spin angular momentum of the black hole). More detailed
simulations and recent observations provide a broad range of values~\citep{2005ApJ...632.1035O}
with a maximum observed spin $(a/M) > 0.98$~\citep{McClintock:2006xd}.  The maximum
spin that a neutron star can have is estimated from a combination of
simulations and observations of pulsar periods. Numerical simulations
of rapidly spinning neutron stars give \checked{$(a/M) <
0.75$}~\citep{Cook:1993qr}; the maximal spin of the observed pulsar
sample may be substantially lower than that.  In our spinning simulations, we
adopted a distribution in which the spin magnitudes are uniformly
distributed between zero and $(a/M) = (cS/GM^2) = 0.98$ and $(a/M) =
(cS/GM^2) = 0.75$ for the black holes and neutron stars respectively,
while the direction of each spin is uniform over the sphere.
There is strong evidence that short GRBs are beamed \citep[see, e.g.,][and references therein]{2006ApJ...650..261S, NakarReviewArticle2006, 2006ApJ...653..468B}, although probably less beamed than long bursts \citep{2006ApJ...653..462G}.
If this is the case, the most likely direction for beaming is along the total
angular momentum vector of the system. For binaries with small
component spins, this will correspond to the direction orthogonal to
the plane of the orbit. Hence the inclination angle of the binary,
relative to the line of sight, is most likely to be close to zero.
However, since zero inclination is the best case for
detection of gravitational waves, a uniform
prior on $\cos \iota$ provides a conservative constraint.
We drew $\cos \iota$ from a uniform prior.

\begin{figure}[h]
\begin{center}
\includegraphics[width=\columnwidth]{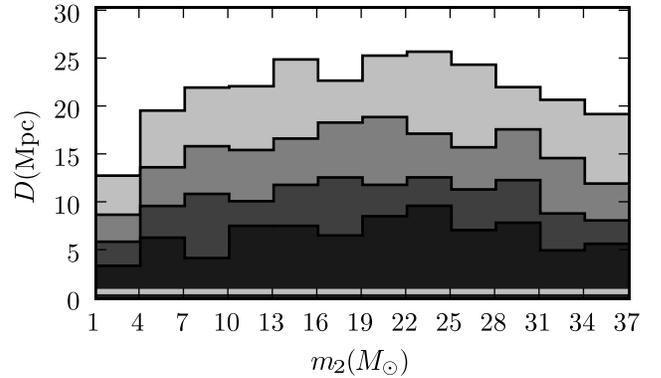}
\caption{\label{f:pcsignal-loudest}
The probability as described in Eq.~(\ref{inspiralProb}) is computed
using injections made only into the 180~s segments immediately before
and after the on-source time. The shaded regions represent $90\%$, $75\%$, $50\%$,
and $25\%$ exclusion regions, from darkest to lightest respectively.
The distance to M31 is indicated by the horizontal line at $D=0.77
\textrm{ Mpc}$. Both amplitude calibration uncertainty and Monte-Carlo
statistics are included in this result; apparent fluctuations as
a function of mass are due to Monte Carlo uncertainty.}
\end{center}
\end{figure}

Figure~\ref{f:pcsignal-loudest} shows the contours of constant probability
$1 - p[0 | \, h(t; m_2,
D)]$.  Compact binaries corresponding to parameters $(m_2, D)$ in the darkest-shaded region
are excluded as progenitors for this event at the $90\%$ confidence
level. As a reference point, a compact binary progenitor with masses
$1~M_\odot < m_1 < 3~M_\odot$ and $1~M_\odot < m_2 < 4~M_\odot$ with $D < \checked{3.5}$~Mpc is
excluded at $90\%$ confidence; the same system with $D < \checked{8.8} \textrm{ Mpc}$ is excluded at the 50\% level.
This result is averaged over different theoretical waveform families;
$20\%$ of the simulated waveforms include spins
sampled as described above.

A number of systematic uncertainties enter into this analysis, but
amplitude calibration error ($\approx 10\%$) and Monte-Carlo statistics
have the largest effects. These uncertainties have been folded into our
analysis in a manner similar to that described in
\citep{LIGO-Inspiral-s2-bns,LIGOS2macho}.  In particular, the
amplitude calibration was taken into account by scaling the distance of the
injection signal to be $1.28 \times 10\%$ larger; the Monte-Carlo
error adds $1.28 \sqrt{p (1-p)/n}$ to $p=p[0 | \, h(t; m_2, D)]$ where
$n$ is the total number of simulated signals in a particular
mass-distance bin.

We evaluate the hypothesis that the event occurred in M31, as electromagnetic
observations hint might be the case, given our observation. We adopt
the measured distance of $0.77 \textrm{ Mpc}$ to M31.
We then simulated a large number of inspirals at distances $0.77
\textrm{ Mpc} < D < 0.9 \textrm{ Mpc}$ which allows us to account for
both uncertainty in distance to M31 ($7\%$)~\citep{2001ApJ...553...47F} and the amplitude calibration
uncertainty discussed above. The simulations exclude any compact binary
progenitor in our simulation space at the distance of M31 at the $>99\%$ level.

\section{Search for a gravitational wave burst}
\label{sec:burstresults}

To search for a gravitational wave burst associated with GRB~070201
we have used LIGO's current baseline method for near-real time searches
for gravitational wave bursts associated with GRB triggers~\citep{GCN,IPN}.
A detailed description of the analysis method is presented elsewhere~\citep{S2-S3-S4_GRB}.

\subsection{Search Method}
\label{ssec:burstmethods}

The burst search method is based on cross-correlating a pair of pre-conditioned
datastreams from two different gravitational wave detectors. The pre-conditioning of the
datastreams consists of whitening, phase-calibration,
and bandpassing from 40 Hz to 2000 Hz.
The cross-correlation is calculated for short time series of equal length
taken from the datastreams of each detector.
For discretely sampled time series $s_1$ and $s_2$, each containing
$n$ elements, the cross-correlation, $cc$, is defined as:
\begin{equation}
cc = \frac{\displaystyle \sum_{i=1}^{n} [s_1(i)-\mu_1][s_2(i)-\mu_2]} {\sqrt[]
          {\displaystyle \sum_{j=1}^{n} [s_1(j)-\mu_1]^2}\; \sqrt[]
          {\displaystyle \sum_{k=1}^{n} [s_2(k)-\mu_2]^2}}
\label{eq:ccdef}
\end{equation}
where $\mu_1$ and $\mu_2$ are the corresponding means of $s_1$ and $s_2$.
Possible values of this normalized cross-correlation range from -1 to +1, the
minus sign corresponding to anti-correlation and the plus sign to correlation.

The measurement of the cross-correlation statistic proceeded as follows. Both
180~s, on-source time series of H1 and H2 data were divided into time intervals (or cross-correlation windows)
of length $T_{ccw}$. Previous analyses have shown that using two
windows,
$T_{ccw}=25\ \mathrm{ms}$ and $T_{ccw}=100\ \mathrm{ms}$, is sufficient to
target short-duration signals lasting from $\sim$~1~ms to
$\sim$~100~ms.
The intervals were overlapped by half (i.e., $T_{ccw}/2$) to avoid missing a
signal occurring near a boundary. The cross-correlation value, $cc$, was calculated for each H1-H2
interval pair and for both $T_{ccw}$ cross-correlation window lengths. The largest $cc$ is the strength measure of the most significant correlated candidate value within the 180~s long on-source segment. To estimate the significance of this loudest event, we use off-source data to measure the cross-correlation distribution of the background noise.

\subsection{Background Estimation and Search Results}
\label{ssec:burst-background}

Approximately 3 hours of data symmetrically distributed about
the on-source segment were used to study the background. These off-source
data were collected from sufficiently close to the
on-source time to accurately reflect the statistical properties of the data within the
on-source region. The detectors were collecting data continuously during the off- and on-source periods. The off-source data were divided into 180~s long segments, corresponding to the length of the on-source segment. The off-source segments were treated identically to the on-source segment.

The distribution of largest $cc$ values in the absence of a signal was
estimated for each cross-correlation window ($T_{ccw}=25\ \mathrm{ms}$ and $T_{ccw}=100\
\mathrm{ms}$) by applying the method in Sec.~\ref{ssec:burstmethods} for all 180~s long off-source
data segments.
To increase the off-source distribution statistics, time shifts between the H1 and H2 datastreams were also performed.
The H1 datastream was shifted by multiples of 180~s
relative to H2.
Then two 180~s stretches from the two detectors were paired at
each shift, making sure that two 180~s time stretches were paired only once.
The distribution of cross-correlations obtained with time-shifted data
is consistent with what is obtained from unshifted data.
For both cross-correlation windows (T$_{ccw}$),
the resulting off-source loudest event $cc$ distribution was used to estimate the probability that background noise alone (i.e.,
without a gravitational wave signal) would produce a $cc$ value larger than the largest
cross-correlation found in the on-source segment.

\begin{figure}
\includegraphics[width=\columnwidth]{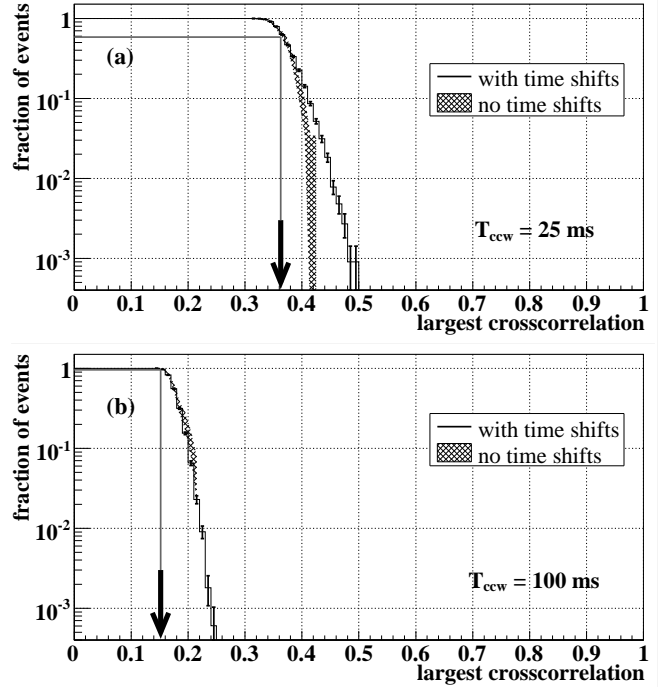}
\caption{\label{fig:posttrials}
Cumulative distribution of measured cross-correlation values
for the (a.) T$_{ccw}=25$~ms and (b.) T$_{ccw}=100$~ms cross-correlation windows.
Both distributions with and without time shifts are shown,
including the statistical errors. The arrows in both cases point to the largest
cross-correlation found in the on-source segment. For the background distributions (1$-$cumulative distribution) is plotted.}
\end{figure}

Figure \ref{fig:posttrials} shows the cumulative cross-correlation
distribution for the $T_{ccw}=25$~ms and $T_{ccw}=100$~ms cases. For the
$T_{ccw}=25\ \mathrm{ms}$ time-window, the largest cross-correlation found in the
on-source data was $cc=0.36$ (see arrow on Figure \ref{fig:posttrials}-a).
The probability of obtaining a cross-correlation value this large from
noise alone is 0.58. For the $T_{ccw}=100\ \mathrm{ms}$ time-window, the
largest cross-correlation found in the on-source data was $cc=0.15$ (see arrow on Figure \ref{fig:posttrials}-b), and the
probability for this cross-correlation value is 0.96.
These results are, therefore, consistent with noise. We conclude that no gravitational wave burst associated with GRB~070201 was detected by the search.

\subsection{Upper Limits on the Amplitude and Energy of Gravitational-Wave Transients Associated with GRB~070201}
\label{sec:burstul}

Since the analysis of the previous section showed no evidence for a
gravitational wave burst, we set upper limits on the amplitude and
energy of gravitational waves incident on the detectors during
GRB~070201. Denote the gravitational wave signal by
$h(t; h_{\mathrm{rss}})$, where
\begin{equation}
h_{\mathrm{rss}} = \sqrt{\int_{-\infty}^\infty \bigl( \, |h_+(t)|^2 + |h_\times(t)|^2 \, \bigr) ~ dt}
\label{eq:hrss}
\end{equation}
is the root-sum-squared amplitude of the gravitational wave signal.
To determine an upper limit, one needs the probability of measuring
$cc$ given the presence of a signal
with $h_{\mathrm{rss}}$:
\begin{equation}
\label{eq:maxcc}
p[cc| h(t; h_{\mathrm{rss}})] \; .
\end{equation}

The search targets signals with duration
$\lesssim 100 \textrm{ ms}$. Within this class of signals, the
sensitivity of the search has weak dependence on signal morphology; it
depends primarily on the energy and the frequency content of the
signal. Therefore, as long as the frequency and duration of the injected test waveforms match the theoretical predictions, we can work with the waveform of our choice. A class of waveforms
called \textit{sine-Gaussians} have become the standard benchmark
for burst searches
and were used to construct the
probability distribution given in Eq.~(\ref{eq:maxcc}). The explicit
formulae for $h_+(t)$ and $h_\times(t)$ are
\begin{eqnarray}
h_+(t) &=\, h_{0}\, \sin(2 \pi f_0 (t-t_0))\, \exp\biggl[\dfrac{-(2\pi f_0 (t-t_0))^2}{2Q^2}\biggr] \,, \label{eq:sg} \\
h_{\times}(t) &=\, h_{0}\, \cos(2 \pi f_0 (t-t_0))\, \exp\biggl[\dfrac{-(2\pi f_0 (t-t_0))^2}{2Q^2}\biggr] \,,
\label{eq:cg}
\end{eqnarray}
where $f_0$ is the central frequency, $h_{0}$ is the peak amplitude of
each polarization, $t_0$ is the peak time, and $Q$ is a dimensionless constant that
represents roughly the number of cycles with which the waveform
oscillates with more than half of the peak amplitude.
Since the
$h_+(t)$ and $h_\times(t)$ waveforms have the same amplitude, these
simulated gravitational wave bursts are circularly polarized.

We provide results for the characteristic case of $Q=8.9$, as the dependence of the upper limits on Q is very weak.
The measurement is carried out as follows.
First, we choose a central frequency, $f_0$, and an $h_{\rm rss}$ value
for the injected signal.  From these parameters, we calculate $h(t)$
using Eq.~(\ref{eq:hresponse}), Eq.~(\ref{eq:hrss}), Eq.~(\ref{eq:sg}) and Eq.~(\ref{eq:cg}).
We then add the calibrated $h(t)$ to the on-source H1 and H2 data, choosing a random starting
time within the segments.
We then measure the largest value of cross-correlation, $cc$, generally following
the same method described in Sec.~\ref{ssec:burstmethods}, except that in this case only a shorter interval around the injection is searched.
Using the same $h_{\rm rss}$ values, we keep iterating the last two steps of the algorithm (randomizing a starting point and calculating the $cc$ local maximum) until we have enough datapoints to determine the conditional probability $p(cc|h_{\rm rss})$.
This probability, determined for different $h_{\rm rss}$ values and central frequencies, is then used to set a frequentist upper limit on $h_{\rm rss}$, given the largest cross-correlation
found for the on-source segment in the search (see Sec.~\ref{ssec:burstmethods})~\citep{S2-S3-S4_GRB}.

The resulting 90\% $h_{\rm rss}$ upper limits are given in Table~\ref{tab:ulsg}
for circularly polarized sine-gaussians with different central frequencies and
with Q = 8.9. The frequency dependence of the upper limits follows the shape of the detector's frequency dependent noise spectrum.

The $h_{\rm rss}$ limits given in Table~\ref{tab:ulsg} include the calibration and statistical errors. These errors were propagated into the 90\% $h_{\rm rss}$ upper limits using the same procedure used in~\citep{S2-S3-S4_GRB}. The $1 \sigma$ errors considered were: (a.) calibration response phase error (10$^{\circ}$); (b.) calibration response amplitude error (10\%); and (c.) statistical error determined through Monte-Carlo simulations (2.1\%).

\begin{table}
\caption{\label{tab:ulsg}
90\% amplitude upper limits and corresponding characteristic energies from
sine-gaussian (SG) waveform simulations assuming 770~kpc as source distance. The $h_{\rm rss}$ limits given in the table already include the calibration and statistical errors~\citep{S2-S3-S4_GRB}.
}
\begin{ruledtabular}
\begin{tabular}{ccccc}
SG central            & T$_{CCW}$ & 90\% UL on             & Characteristic                           &  Characteristic      \\
frequency (Hz)        & ms        & $h_{\rm rss}$ (Hz$^{-1/2}$) & $E_{\rm GW}^{\rm iso}$ ($M_{\odot}c^2$)  &  $E_{\rm GW}^{\rm iso}$ (erg) \\ \hline
\hphantom{0}100       & 25        &  $2.15\times10^{-21}$  &  $5.8\times10^{-4}$                        &  $1.0\times10^{51}$  \\
\hphantom{0}150       & 25        &  $1.27\times10^{-21}$  &  $4.6\times10^{-4}$                        &  $8.2\times10^{50}$  \\
\hphantom{0}250       & 25        &  $1.34\times10^{-21}$  &  $1.4\times10^{-3}$                        &  $2.5\times10^{51}$  \\
\hphantom{0}554       & 25        &  $2.36\times10^{-21}$  &  $2.1\times10^{-2}$                        &  $3.8\times10^{52}$  \\
\hphantom{0}1000      & 25        &  $4.12\times10^{-21}$  &  $2.1\times10^{-1}$                        &  $3.8\times10^{53}$  \\
\hphantom{0}1850      & 25        &  $7.56\times10^{-21}$  &  $2.5$                                     &  $4.5\times10^{54}$  \\ \hline
\hphantom{0}100       & 100       &  $1.97\times10^{-21}$  &  $4.9\times10^{-4}$                        &  $8.8\times10^{50}$  \\
\hphantom{0}150       & 100       &  $1.25\times10^{-21}$  &  $4.4\times10^{-4}$                        &  $7.9\times10^{50}$  \\
\hphantom{0}250       & 100       &  $1.41\times10^{-21}$  &  $1.6\times10^{-3}$                        &  $2.9\times10^{51}$  \\
\hphantom{0}554       & 100       &  $2.52\times10^{-21}$  &  $2.5\times10^{-2}$                        &  $4.5\times10^{52}$  \\
\hphantom{0}1000      & 100       &  $4.51\times10^{-21}$  &  $2.6\times10^{-1}$                        &  $4.7\times10^{53}$  \\
\hphantom{0}1850      & 100       &  $8.15\times10^{-21}$  &  $2.9$                                     &  $5.2\times10^{54}$  \\
\end{tabular}
\end{ruledtabular}
\end{table}

The upper limits on $h_{rss}$ implied by the burst search can be translated
into conventional astrophysical units of energy emitted in
gravitational waves.  The gravitational wave energy $E_{\rm GW}$ radiated
by an {\it isotropically emitting} source that is
dominated by emission at a frequency $f_0$, is related to the
$h_{\rm rss}$ received at distance $D$, much less than the Hubble distance, by~\citep{EnergyFormula}
\begin{equation}
E_{\rm GW}^{\rm iso} \approx \frac{\pi^2 c^3}{G}D^2 f_0^2 \, h_{\rm rss}^2 \, .
\label{eq:Emerge}\end{equation}

Based on the sensitivity of this burst search as summarized in Table~\ref{tab:ulsg}, we estimate that a gravitational wave burst with characteristic frequency in the most sensitive frequency region of the LIGO detectors ($f \approx 150\,\text{Hz}$), if GRB~070201 originated in M31 (at 770~kpc), must have emitted less than approximately 4.4$\times$10$^{-4} M_{\odot}c^2$ (7.9$\times$10$^{50}$ ergs) within any 100 ms interval in the on-source window in gravitational waves.
In terms of the SGR progenitor hypothesis, our experimental upper limit on $E_{\mathrm{GW}}$ is several orders of
magnitude larger than the $10^{45}\,\text{erg} \, (D/770 \, {\rm kpc})^2$ known
to be emitted electromagnetically. And while present models for SGR
bursts may differ substantially in their mechanism~\citep{pacheco98,2001MNRAS.327..639I,Owen:2005fn,2001MNRAS.327..639I,horvath05},
they suggest that no more than $10^{46}\,\text{erg}$ is released in the form of gravitational waves.
Therefore, the upper limit achievable with the present detectors does not exclude
these models of SGRs at the M31 distance.

\begin{figure}
\includegraphics[width=\columnwidth]{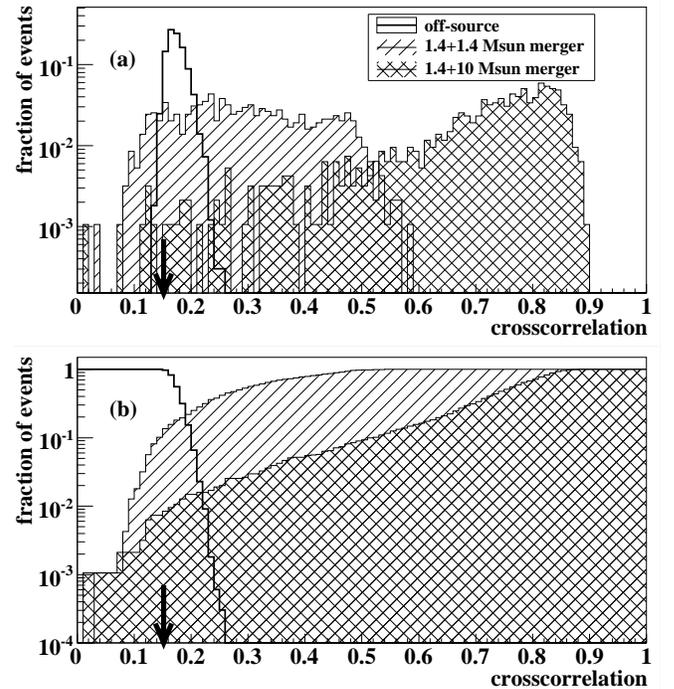}
\caption{\label{fig:BI-posttrials}
Differential (a.) and cumulative (b.) distributions of measured cross-correlation values
for the T$_{ccw}=100 \,\mathrm{ms}$ cross-correlation window.
Distributions for both the $1.4M_\odot$--$1.4M_\odot$ and $1.4M_\odot$--$10M_\odot$ binaries are shown along with the background distribution. The arrows points to the largest
cross-correlation found in the on-source segment. On plot (b.) (1$-$cumulative distribution) of the off-source data is plotted. }
\end{figure}

We also estimate the sensitivity of the (100~ms) burst search to gravitational waves from
a compact binary progenitor in M31 (see Figure~\ref{fig:BI-posttrials}).  We choose as examples
a $1.4M_\odot$--$1.4M_\odot$ binary and a $1.4M_\odot$--$10M_\odot$ binary.
For each mass pair, we inject approximately 1000 inspiral waveforms consistent
with the distance of M31, with random isotropically distributed inclination
and polarization, and with coalescence time uniformly distributed through the
on-source segment.  Since, for these masses, the merger phase is expected to
occur at frequencies well above that of maximum LIGO sensitivity, we inject
only the inspiral portion. As for the sine-Gaussian simulations, we determine
the largest cross-correlation within a small time window around the coalescence
time.  We also account for possible systematic errors due to the calibration and
the uncertainty in the distance to a possible source within M31, and statistical
errors from the Monte Carlo procedure.  We estimate with 90\% confidence that
a $1.4M_\odot$--$1.4M_\odot$ binary inspiral in M31 at the time of GRB~070201
would have a probability of at least 0.878 of producing a cross-correlation larger
than the loudest on-source event.
For $1.4M_\odot$--$10M_\odot$ binaries this probability is at least 0.989.
This gives us an independent way to reject the hypothesis of a compact
binary progenitor in M31, while not relying on the detailed model of
the inspiral signal.

\section{Discussion}

We analyzed the data from the LIGO H1 and
H2 gravitational-wave detectors, looking for
signals associated with the electromagnetic event GRB~070201.
No plausible gravitational-wave signals were identified.
Based on this search, a compact binary
progenitor (neutron star + black hole or neutron star systems) of GRB~070201, with masses in the range $1~M_\odot < m_1 <
3~M_\odot$ and $1~M_\odot < m_2 < 40~M_\odot$, located in M31 is
excluded at the $> 99\%$ confidence.

Our model-independent search did not find correlated signatures inconsistent with the noise within the H1 and H2 data-streams that could be related to GRB~070201. Based on the sensitivity of our search and assuming isotropic gravitational-wave emission of the progenitor, an upper limit on the power emitted in gravitational waves by GRB~070201 was determined. A gravitational wave with characteristic frequency within the most sensitive range of the LIGO detectors ($f\approx$~150~Hz) \checked{most probably} emitted less than
E$_\mathrm{GW} <$7.9$\times$ 10$^{50}$~ergs within any 100-ms-long time interval inside the on-source region if the source is in M31. This limit on radiated power is comparable to the emitted power of some GRBs. However, it is significantly higher than the associated electromagnetic emission of this particular GRB. Therefore the transient search only constrains the binary inspiral models for a short hard GRB in M31 and does not impose new limitations on magnetar-driven (SGR type) burst models.

As gravitational-wave observations continue and the sensitivity of the
instruments improves, we look forward to the astrophysical
insights that combined electromagnetic and gravitational observing
campaigns can bring.

\acknowledgments
We are indebted to the observers of the electromagnetic event, GCN and IPN for providing us with valuable data and real time information.
We are grateful to Neil Gehrels of NASA/GSFC for his help in reviewing the article.
The authors gratefully acknowledge the support of the United States
National Science Foundation for the construction and operation of the
LIGO Laboratory and the Science and Technology Facilities Council of the
United Kingdom, the Max-Planck-Society, and the State of
Niedersachsen/Germany for support of the construction and operation of
the GEO-600 detector. The authors also gratefully acknowledge the support
of the research by these agencies and by the Australian Research Council,
the Council of Scientific and Industrial Research of India, the Istituto
Nazionale di Fisica Nucleare of Italy, the Spanish Ministerio de
Educaci\'on y Ciencia, the Conselleria d'Economia, Hisenda i Innovaci\'o of
the Govern de les Illes Balears, the Scottish Funding Council, the
Scottish Universities Physics Alliance, The National Aeronautics and
Space Administration, the Carnegie Trust, the Leverhulme Trust, the David
and Lucile Packard Foundation, the Research Corporation, and the Alfred
P. Sloan Foundation.
We are grateful to the GALEX collaboration for providing the UV image of M31, the SDSS project for providing the optical image of M31 and Google Sky to accelerate the error box position verification.
This paper was assigned a LIGO Document Number: LIGO-\ligodoc.

\bibliographystyle{apj}

\end{document}